\documentclass[preprint,12pt]{elsarticle}

\usepackage{graphicx}
\usepackage{amssymb}
\usepackage{widetable}
\usepackage{url}

\journal{submitted to...}

\begin{document}
	
	\begin{frontmatter}
		
	
		\title{Multiple Factor Analysis of Distributional Data}
		\author{Rosanna Verde, Antonio Irpino}
		\address{rosanna.verde@unina2.it, antonio.irpino@unina2.it \\
			Dept. of Mathematics and Physics \\
			University of Campania L. Vanvitelli\\
			Viale Lincoln 5, 81100, Caserta, Italy}


		\begin{abstract}
			In the framework of Symbolic Data Analysis (SDA), distribution-variables are a particular case of multi-valued variables: each unit is represented by a set of distributions (e.g. histograms, density functions or quantile functions), one for each variable.
			Factor analysis (FA) methods are primary exploratory tools for dimension reduction and visualization.
			In the present work, we use Multiple Factor Analysis (MFA) approach for the analysis of data described by distributional variables.
			Each distributional variable induces a set new numeric variable related to the quantiles of each distribution. We call these new variables as \textit{quantile variables} and the set of quantile variables related to a distributional one is a block in the MFA approach. Thus, MFA is performed on juxtaposed tables of quantile variables. \\
			We show that the criterion decomposed in the analysis is an approximation of the variability based on a suitable metrics between distributions: the squared $L_2$ Wasserstein distance. \\
			Applications on simulated and real distributional data corroborate the method. The interpretation of the results on the factorial planes is performed by new interpretative tools that are related to the several characteristics of the distributions (location, scale and shape).
		\end{abstract}
		
		
	\end{frontmatter}
	
	
\section{INTRODUCTION}
\noindent
In the framework of Symbolic Data Analysis (SDA) \emph{distributional} (or distribution-valued) data  (DD) are data described by distributions (like histograms, density or quantile functions) for a numeric variable. Such data are realizations of a numeric \emph{modal} symbolic variable, according to \cite{Bo_did}, also defined as \emph{distributional} variable.
In general, a distributional data can be expressed by a parametric or a non-parametric density function estimated on a set of observed values.

Techniques of dimension reduction have been extended to the analysis of multi-valued variable in order to visualize the proximity between the individuals and the correlations between the variables onto lower dimensional spaces.
Factorial methods, like Principal Component Analysis (PCA), are the most usual techniques for a dimensional reduction of a set of $p$ numeric variables observed on $n$ individuals.
They aim at extracting a set of new \emph{orthogonal} factors, which explain the variance-covariance structure through few linear combinations of the original variables.

As for all other factorial techniques, PCA reduces the redundancy of information presents in the data such that the more the variables are correlated, the higher is the dimensionality reduction.

When data are distributions, the extraction of new factorial axes should take into account the characteristics of distributions as well as the variability among distributions.
While the meanings of orthogonality, variance, covariance and correlation are consolidated for classical numeric variables, they are not for distributional ones.

In the framework of Symbolic Data Analysis, different PCA methods have been proposed for interval-valued data (see \cite{PCA_wiley} and \cite{GCA_wiley} for an extensive review), while few proposals exist for distributional ones.

Some proposals have been designed for histogram-valued data \cite{histo_PCA1,histo_PCA3, histo_PCA2, Lerad_08, hitoPCA6} and they substantially differ for the variability criterion decomposed in the analysis.

\cite{histo_PCA1} proposed an extension of a PCA for interval data \cite{cPCA} to histogram data by considering intervals of relative frequencies. In such approach, it is supposed that the histograms share a common partition of the support (i.e. the same set of bins) and the analysis is conducted only on a transformation of the frequencies of the bins of histogram data.
The decomposed covariance structure of the data takes into account only the covariance of the centers of the intervals of frequencies, while the information related to the numerical support of the histograms is ignored.
In a second contribution \cite{histo_PCA3}, a the variance-covariance matrix of a set of multivariate distributions is decomposed under the hypothesis of conditional independence. In this case, conditional independence assumption  leads to consider only the covariances between the means,  and the variability related to different size and shape of the distributions is lost.
Another proposal \cite{histo_PCA2} consists in a PCA for histogram-valued data considering only the empirical frequencies observed for each bin of the observed histograms, losing, like in \cite{histo_PCA1} the information related to the support.
\cite{histo_PCA4, histo_PCA5} proposed a PCA of \emph{quantile representations} of symbolic data (they are particular transformations of the observed distributional data). However, the author does not define the geometric properties of the decomposed covariances explicitly, as well as he does not give a suitable  interpretation for the explained variability on the factorial sub-spaces.
\cite{Lerad_08} proposed an extension of the interval PCA to histogram-valued data (considered as weighted intervals). In this case, the eigenvalues of the PCA decompose an inertia measure that correspond to the sum of the variances of histogram-valued variables as presented in \cite{BillDid06}.

Finally,  an extension of the interval PCA to histogram variables is proposed by \cite{hitoPCA6}, where a PCA is performed on the means of the histograms (similarly than the centers PCA). Then, using the Tchebicheff inequality, the histograms are transformed into intervals and projected on the space spanned by the principal components. Unfortunately, the principal components are related only to the covariances of the means of the histograms. More recently, \cite{CYB1_PCA} proposed an adaptation of the previous methods for the PCA of normal-distribution-valued data.

All the above mentioned methods do not require  explicitly the definition of a measure of covariance between distributional variables in advance.
Some basic statistics for histogram variables were presented by \cite{Bo_did} and developed by \cite{BillDid06}. Recently, \cite{Noi1} proposed new variance, covariance and correlation measures for distributional variables based on the  $\ell_2$ Wasserstein distance \cite{Rush01} between distributions.

Both the approaches confirm that the variability of a distributional variable can be decomposed in several components: in the approach of \cite{BillDid06}, the data variability is expressed in a part related to the location and in a part related to scale; while, in the approach of \cite{Noi1}, it is taken into consideration also the shape of the observed distributions.
In particular, the results in \cite{Noi1}, consistently with the statistical modeling of quantile functions proposed by \cite{Gilch00}, show that the analysis conducted on empirical quantile functions (the inverse of the cumulative distribution functions) has two main interpretative advantages. Firstly, working directly on the quantile functions associated with the empirical distributions, it is not necessary to consider parametric hypothesis about the distributions. Secondly, it is possible to interpret the contribution to the results related to variability of the location, the scale and the shape of the distributions separately.
More recently \cite{IEEE_PCA} proposed a PCA method for a single distribution variable using an approximation of the Wasserstein distance between distributions. The idea is to represent the distributional variables through a set of quantile variables and then to apply the PCA on a quantile matrix. The data are not standardized, so that the information about the several characteristics: location, size and shape of the distributions is retrieved in the determination of the new factorial axes. The results of the analysis furnish an interesting interpretation of the axes accordingly to the several moments of the distributions.

The present paper aims at using Multiple Factor Analysis (MFA) to analyse data described by a set of distributional variables.
Multiple Factor Analysis (MFA) has been introduced by the works of \cite{MFA83,MFA83_1,MFA90} (recent overviews are available in \cite{MFA-CS13} and \cite{Pages_14}). MFA is an extension of principal component analysis (PCA) on several  sets of variables (namely, blocks of variables) and it is part of multi-tables techniques (e.g., STATIS, Multiblock Correspondence Analysis, SUM-PCA). MFA proceeds in two steps: first, it runs a PCA of each block of variables, then it normalizes each block by the respective first singular value, so that their first principal components have the same length. Second, it performs a common representation of the data sets that is called compromise, or consensus representation. This compromise is obtained from a (non-normalized) PCA of a table obtained from the concatenation of the normalized blocks of variables.

Here, we propose Multiple Factor Analysis (MFA) on a transformation of distributional variables in sets of quantile variables, one for each variables, where each set of quantile variables related to a distributional one is a block of variables in the MFA. The peculiarity of this approach is to decompose an approximation of the total variability of the distributional data according to the squared $\ell_2$ Wasserstein distance (see \cite{IEEE_PCA}). In this way, we preserve the coherence between the criterion optimized in MFA and the distributional data space defined by the $\ell_2$ Wasserstein metric.
Finally, the proposal carries out visualization and graphical interpretative tools to analyze the relationships between distributions according to their own characteristics: location, scale and shape, on factorial planes.

The remainder of paper is structured as follows. Section \ref{sec:2} introduces the data and the Wasserstein  metric between distributions. Section \ref{sec:3} presents the extension of the  Multiple Factor Analysis on quantile variables with the aim of analyzing the relationships between more distributional variables observed on the same set of individuals. Section \ref{sec:4} shows a new tool for the visualization of the distributional variables on the factorial planes. Sections \ref{sec:5} and \ref{sec:6} shows the results of an application of the proposed method on simulated and real data, respectively. Finally, Section \ref{sec:conl} concludes the paper.
\vspace{2pt}

\section{DISTRIBUTIONAL VARIABLES AND THE WASSERSTEIN DISTANCE}\label{sec:2}
\noindent
Let $E$ be a set of $n$ individuals described by a distributional variable $Y$, i.e.,  a modal-valued variable with numerical domain $\mathcal{S}=[min(Y), Max(Y)] \subset \Re$.
We denote with $y_i$ ($i=1,...,n$), the realization of the variable $Y$ for the $i-th$ individual \cite{Noi1}, expressed by an (empirical or estimated) probability  distribution  $f_i(y)$. We denote with $F_i(y)$ the cumulative density function (cdf) and with $F^{-1}_i(t)$ (for $t\in[0;1]$) the quantile function (qf, i.e. the inverse of the cdf). According to \cite{Gilch00}, several advantages can arise by working with \emph{qf}s rather than with the distribution functions: all the \emph{qf}s have a finite domain in $[0;1]$, the sum of quantile functions returns a \emph{qf}, the product of a \emph{qf} by a positive scalar returns a \emph{qf}, under certain conditions it is possible to define the product between two \emph{qf}s.

Several proposals have been formulated in the framework of Symbolic Data Analysis (SDA) to define univariate (mean, variance, standard deviation) and bivariate (covariance, correlation) statistics for histogram variables \cite{Bo_did, BillDid06}. Recently, \cite{Noi1} have introduced new measures based on the Wasserstein distance, that is a suitable metric to compare distributions.
An overview of the family of Wasserstein metrics is in \cite{Rush01, Villani03}.

According to \cite{Rush01},  the $\ell_p$ Wasserstein distance between two (univariate) distribution functions can be expressed as:

\begin{equation}\label{form_WassDist}
d_{W_p}(y_i,y_{i'})=  \left[\int\limits_0^1 {\left| {F _i ^{ - 1}(t) - F_{i'} ^{ - 1} (t)}\right|^p dt}\right]^\frac{1}{p}
\end{equation}

where, $p\geq1$, $F_i$ and $F_{i'}$ are cumulative distribution functions (\emph{cdf}s) associated with the $y_i$ and $y_{i'}$ histograms and $F_i ^{-1}$ and $F_{i'}^{-1}$  the corresponding quantile functions (\textsl{qf}s).
The $\ell_2$ squared Wasserstein distance, also known as Mallow distance \cite{Rush01}, between the {\em qsf} associated with  two histograms is:
\begin{equation}\label{HOMSQ_1}
d^2_{W_2} (y_i,y_{i'})=\int\limits_{0}^{1} {\left[{F_i^{-1}(t)-F_{i'}^{-1}(t)} \right] ^2 dt}.
\end{equation}
The $\ell_2$ Wasserstein metric can be considered as a natural extension of the Euclidean metric between quantile functions.

Thus, the Wasserstein distance can be suitably computed for equi-depth histograms with a fixed number of bins equal to $s$. Given a histogram description $y_i$, partitioned into $s\geq1$ bins:
$$ y_i= \left\{ {\left( {I_{1i}},\pi _{1i}  \right),...,\left( {I_{hi}} ,\pi _{hi}
	\right),...,\left( I_{s i} ,\pi _{s i} \right)} \right\},$$
where  $I_{hi}=[\underline {y}_{hi};\overline {y}_{hi}]$ is an interval of $\Re$, and $\pi_{hi}\geq 0$ such that $\sum_{h=1}^{s}\pi_{hi}=1$.
If $y_i$ is an \emph{equi-depth histogram} then $\pi_{hi}=\frac{1}{s}$.
The following quantities $w_{li}$ represent
the cumulative weights associated with the elementary intervals of
$y(i)$:
\begin{equation}\label{distribwhei}
w_{li}  = \left\{ {\begin{array}{*{20}c}
	0 & {l = 0}  \\
	{\sum\limits_{h = 1, \ldots ,l} {\pi _{hi} } } & {l = 1, \ldots ,s }  \\
	\end{array}\hspace{0.5cm}. } \right.
\end{equation}

For simplicity, we consider two equi-depth histogram descriptions
$y(i)$ and $y(i')$ having the same number of bins equal to $s$, it
implies that the weights $\pi_{li}=\pi_{li'}=w_{li}-w_{l-1i}=\frac{1}{s}$. In this case, being $w_{li}=w_{li'}$, we omit the second index.
The squared Wasserstein distance between two equi-depth histogram descriptions is computed as:
\begin{equation}\label{HOMSQ}
d^2_{W_2}(y_i,y_{i'}): =  \sum\limits_{l=1}^s {\int\limits_{w_{l-1}}^{w_l} {\left( {F_i^{ - 1} (t) - F_{i'}^{ - 1}
			(t)} \right)^2dt} }.
\end{equation}

Each couple $(w_{l-1},w_l)$ allows us to identify two uniformly
dense intervals, one for $i$ and one for $i'$, having respectively
the following bounds:
\begin{displaymath}
\begin{array}{l}
I_{li}=[F_{i}^{-1}(w_{l-1});F_{i}^{-1}(w_{l})]=[\underline{y}_{li};\overline{y}_{li}]
\;\textrm{and}\\
I_{li'}=[F_{i'}^{-1}(w_{l-1});F_{i'}^{-1}(w_{l})]=[\underline{y}_{li'};\overline{y}_{li'}].
\end{array}
\end{displaymath}

The center and the radius of each interval are computed as follows:
\begin{displaymath}
c_{li}=(\underline{y}_{lu}+\overline{y}_{lu})/2
\hspace{10pt}  r_{lu}=(\underline{y}_{lu}+\overline{y}_{lu})/2  \hspace{10pt}   for \hspace{5pt} u=i,i'.
\end{displaymath}

Since the histograms are equi-depth, all the $\pi_l$ are equal to $1/s$. The
intervals are uniformly distributed can be expressed as functions of their centers and radii. Hence, the equation (\ref{HOMSQ}) can be rewritten as follows:
\begin{equation}\label{HOMSQfin}
d^2_{W_2} (y_i,y_{i'}) = \frac{1}{s} \sum\limits_{l=1}^s { \left[{\left({{c_{li}-c_{li'}}}\right)^2 + \frac{1}{3}\left({{r_{li}-r_{li'}}}\right)^2}\right]}.
\end{equation}

According to the Wasserstein metric, the \emph{mean histogram}  $y_b$ is defined as a Fr\'echet mean by solving the following minimization problem:

\begin{equation}\label{dist_b}
f(y_b)=\arg\min_y \sum_{i=1}^n d_{W_2}^2(y_i,y),
\end{equation}

that is expressed as follows:

\setlength{\arraycolsep}{0.0em}
\begin{eqnarray}\label{dist_fb}
y_b=\left[ \left((c_{1b}-r_{1b};c_{1b}+r_{1b}),\frac{1}{s} \right)\right.;\ldots;
\left.\left((c_{sb}-r_{sb};c_{sb}+r_{sb}),\frac{1}{s} \right)\right];
\end{eqnarray}

where:
\begin{displaymath}
c_{lb}  = n^{-1}\sum\limits_{i = 1}^n { c_{li} }  \hspace{10pt};\hspace{10pt}
r_{lb}  = n^{-1}\sum\limits_{i = 1}^n { r_{li} }.
\end{displaymath}

Therefore, the variance of the distributional variable $Y$, according to (\ref{dist_b}) and (\ref{dist_fb}), is defined as follows:

\setlength{\arraycolsep}{0.0em}\begin{eqnarray}
\label{varianceW}
Var(Y)=\frac{1}{n}\sum_{i=1}^n d_{W_2}^2(y_i,y_b) =\frac{1}{n\,s}\sum_{i=1}^n \sum_{l=1}^s  { \left[{\left({{c_{li}-c_{lb}}}\right)^2 + \frac{1}{3}\left({{r_{li}-r_{lb}}}\right)^2}\right]}.
\end{eqnarray}
\setlength{\arraycolsep}{5pt}

In the next section, we will show that $Var(Y)$ can be approximated by the sum of the variances of the quantile variables of the distributional variable $Y$.

If histograms are not equi-depth, the computations are done accordingly to \cite{Noi_adac}. In such a case, it is requested a homogenization step for comparing histograms through the Wasserstein distance. The support of the distributional data is shared according to a set of quantile values corresponding to the same set of density levels $p_i$ for all the distributions.

\subsection{Other properties of the $\ell_2$ Wasserstein distance}

The advantage of using the $\ell_2$ Wasserstein distance for comparing distributional data is related to an important property that it satisfies: the squared distance between two distributions can be decomposed according to the following three components (as proved by \cite{IRPRO}):

\begin{equation}\label{eq:proof}
\begin{array}{l}
d_W^2 (y_i,y_{i'} )
= \int\limits_0^1 {\left[ {F_i^{ - 1} (t) - F_{i'}^{ - 1} (t)} \right]^2 dt}=\\
=\underbrace {\left( {\mu_i - \mu _{i'} }
	\right)^2}_{Location} + \underbrace{\underbrace {\left( {\sigma _i - \sigma _{i'}
		} \right)^2}_{Scale} + \underbrace {2 \sigma_i \sigma_{i'} (1 -
		\rho_{i,i'} )}_{Shape}}_{Variability}.
\end{array}
\end{equation}

where: $\mu_u$ and $\sigma _{u}$ (with $u=i,i'$) are respectively the means and the standard deviations of the distributions $y_i$ and $y_{i'}$, while $\rho_{i,i'} $,
is the Pearson correlation coefficient between two quantile functions $F_i^{-1}(t)$ and $F_{i'}^{-1}(t)$

Therefore, $\rho_{i,i'} $ can be considered as a measure of shape similarity of two distribution functions. In fact, $\rho _{i,i'}=1$ only if the two distributions have the same standardized quantiles (by the respective mean and standard deviation), which occurs when the two distributions have the same shape.\\
The  decomposition of the squared $\ell_2$ Wasserstein distance between two distribution functions allows to evaluate their deviation in terms of {\em Location}, {\em Scale} and {\em Shape}. The difference in {\em Location} and {\em Scale} are respectively expressed by the squared Euclidean distances between the means and between the standard deviations of the two distributions; while the difference in {\em Shape} is related to the value of $\rho_{i,i'}$. The {\em Scale} and {\em Shape} components express together the difference of the \emph{Variability} structure between two distributions.
\vspace{2pt}

\section{MULTIPLE FACTOR ANALYSIS ON THE QUANTILES OF DISTRIBUTIONAL VARIABLES}\label{sec:3}
\noindent

In this section, we present a MFA on a set of data tables containing the quantile representation of several distributional variables observed on the same individuals. MFA is an extension of PCA aiming to provide a set of common factors for projecting data described by blocks of variables onto a compromise subspace \cite{MFA83}.
The main idea is to extend the PCA methods for distributional data \cite{IEEE_PCA} to the case of multi-tables analysis.

According to the Principal Component Analysis (PCA) strategy on a distributional variable $Y$, distributions are replaced by a set of predefined quantiles, which are assumed as values of the variables of analysis.

Let $E$ be the set of $p$ histograms $y_{ij}$ (for $j=1,\ldots, p$) related to the description of the $i-th$ individual w.r.t. the $Y_1, \ldots, Y_j \ldots, Y_p$ variables. Each $y_{ij}$ is the histogram of values that the individual $i-th$ assumes of the variable $Y_j$.
We consider that all the histograms are equi-depth, so that the bounds of the intervals $I_{li}=[\underline {y}_{li}; \overline{y}_{li}]$ (for $l=1, \ldots, K_j -1$), correspond to the $K_j$-quantiles, that is, the values which divide the distribution in $K_j$ equal parts). We have denoted $K_j$ the number of quantiles for each variable $Y_j$, that can be also chosen as different for each of them ($K_1 \ldots K_j \ldots K_p$).

For the generic variable $Y_j$, we denote with:

\begin{eqnarray*}
	q_{i0,j}&=&\underline {y}_{1i,j} = min(y_{ij}), \\
	q_{il,j}&=&\overline {y}_{li,j}\;\;(for\;\; l=1, \ldots, K_j -1)\;\;\mathrm{and} \\
	q_{is,j}&=&\overline {y}_{K_ji,j} = Max(y_{i,j})\;\;\forall i=1, \ldots, n
\end{eqnarray*}

In order to perform a PCA on quantiles, we consider as input a concatenation of classic $n \times (K_j+1)$ data tables (with
$j=1, \ldots, p$), denoted with $ \mathbf{Q}_j $:

\begin{equation}
\mathbf{Q} = \left[\mathbf{Q}_{1} |  ...  |  \mathbf{Q}_{j} | ...   |  \mathbf{Q}_{p} \right]
\end{equation}

Each row of the $ j-th $ table $\mathbf{Q}_j$ is an individual representation expressed by the following order statistics: the minimum value (or $\emph{zero}$ quantile) $q_{i0}$;  the $l$-quantiles $q_{il}$; the maximum value or $K_j-th$ quantile, $q_{iK_j}$.

The generic $i-th$ individual (row) observed for the (single) distributional variable $Y_j$ is described by a set of $(K_j+1)$ quantiles (columns): $Q_{0j}, \ldots,Q_{lj},\ldots,Q_{K_j j}$ with $1/K_j$ the probability, or the relative frequency of the observed values between two consecutive quantiles.

\begin{displaymath}
\mathbf{Q}_j=
\left[
\begin{array}{ccccccc}
& q_{10,j} & q_{11,j} & ... q_{1l,j} & ...  & q_{1K_j,j} \\
& ... & ... & ... & ... & ... \\
& q_{i0,j} & q_{i1,j} & ... q_{il,j} & ...  & q_{iK_j,j} \\
& ... & ... & ... & ... & ...  \\
& q_{n0,j} & q_{n1,j} & ... q_{nl,j} & ...  & q_{nK_j,j} \\
\end{array}
\right]
\end{displaymath}

We assume that the elements of the matrix $\mathbf{Q}_j$ are centered by subtracting the means of the respective quantile variables $Q_{lj}$ (for $l=1, \ldots, K_j$).

For simplicity, we refer to the columns of the matrix $\mathbf{Q}_j$ as centered \emph{quantile-variables}. The choice not to standardize quantile variables preserves the approximation of the variance of a distributional variable based on Wasserstein metric approximated by the sum of the variances of the quantile variables (as show hereafter). A particular care should be taken for the the lower and higher quantile variables. Indeed, the empirical evidence (see applications on simulated and real data) reveals that those quantile variables may have a higher variability with respect to the other ones. This can be checked before the analysis. A practical solution is to consider extremes quantile variables as supplementary in the analysis or to give lower weights with respect to the other ones.

We denote $\bf{W}$ the matrix of the individual weights; assuming that all of them have same weight, it is a diagonal matrix of elements $\frac{1}{n}$.

Moreover, we define the cross-product of quantiles matrix $\mathbf{Q}_j$, weighted by $\bf{W}$, as follows:

\begin{equation}
\mathbf{S_j} = \mathbf{Q}_j^{\sf T} \mathbf{W} \mathbf{Q}_{j}
\end{equation}

The $\mathbf{S}_j$ is the variance-covariance matrix of the quantile variables of $Y_j$.

Then, the cross-product of the matrix $\mathbf{Q}$, weighted by $\bf{W}$ is:

\begin{equation}
\mathbf{S}=\mathbf{Q}^{\sf T} \mathbf{Q} = \left[ \mathbf{Q}_{1} |  ...  |  \mathbf{Q}_{j} | ...   |  \mathbf{Q}_{p} \right]^{\sf T}  \mathbf{W} \left[\mathbf{Q}_{1} |  ...  |  \mathbf{Q}_{j} | ...   |  \mathbf{Q}_{p} \right] = \sum_j \mathbf{Q}_{j}^{\sf T} \mathbf{W} \mathbf{Q}_{j} = \sum_j \mathbf{S}_{j}
\end{equation}

The $\mathbf{S}$ is the block variance-covariance matrix of the quantile variables $Q_{lj}$ (for $l=0, \ldots, s$) of the $Y_1, \ldots, Y_j \ldots, Y_p $.

The trace of the matrix $\mathbf{S}$  (denoted $Tr(\mathbf{S})$) it is equal to the sum of the Variances of the quantile variables $Q_{lj}$ (denoted $Var(Q_{lj})$)  (for $l=0, \ldots, s$ and $j=1, \ldots, p$).

Now we show the relationship between the usual $\ell_2$ Wasserstein metric used in the analysis of distributional data and the criterion decomposed in MFA.

In \cite{IEEE_PCA} is showed that the trace of($\mathbf{S}_{j}$) (denoted $Tr(\mathbf{S}_{j})$) approximates the variances
of the distributional variable $Y_j$ (denoted $Var(Y_j)$), according to the $\ell_2$ Wasserstein distance.

Denoting with $\Delta$ the following deviation:

\begin{equation}\label{Delta_trace}
\Delta= Tr(\mathbf{S}_j)-Var(Y_j).
\end{equation}

it depends on the number of quantiles and by the number $K_j$ (with $K=\sum\limits_{j=1}^p K_j $) of the intervals (bins) of the supports of the $n$ histogram data, as follows:

\begin{displaymath}
\Delta= \frac{\sum\limits_{i=1}^n \sum\limits_{l=0}^{K_j} ({q}^c_{il,j})^2 - \sum_{i=1}^n \sum_{l=1}^{K_j} { \left[{\left({c^c_{il,j}}\right)^2 + \frac{\left(r^c_{il,j}\right)^2}{3}}\right]}}{n \cdot K}.
\end{displaymath}

with: $c^c_{il,j}=c_{il,j}-\bar{c}_{l,j},\; r^c_{il,j}=r_{il,j}-\bar{r}_{l,j}$,
the rescaled center and radius on the respective means.

\subsection{The two Steps of MFA}.

MFA is performed in two steps.

The first step consists in a PCA on each data table $\mathbf{Q}_j$. The results are obtained by the SVD decomposition:

\begin{equation}
\mathbf{Q}_j= \mathbf{U}_j \Lambda_j \mathbf{V}_j^{\sf T}
\end{equation}

under the ortho-normality constraints:

$\mathbf{U}_j^T \mathbf{U}_j = \mathbf{V}_j^T \mathbf{V}_j = \mathbf{I}$.

The factor scores are computed as:
\begin{equation}
\Psi_j=\mathbf{U}_j \Lambda_j
\end{equation}

where $\Lambda_j$ is the diagonal matrix of the eigenvalues of the matrix $\mathbf{Q}_j$.

In MFA each table $\mathbf{Q}_j$ is normalized by the respective first squared eigenvalue $\lambda_{1j}$, corresponding to the highest value of $\Lambda_j$, that is:

\begin{equation}
a_j= \frac{1}{\sigma}
\end{equation}

where: $\sigma=\lambda_{1j}^2$.

The $a_j$ for $j=1, \ldots, p$ can represent a system of weights for each matrix and they can be arranged in a diagonal matrix $\mathbf{A}$ :

\begin{equation}
\mathbf{A} = diag \{ [ a_1 \mathbf{1}_{[K_1]}^{\sf T}, \ldots, a_j \mathbf{1}_{[K_j]}^{\sf T},\ldots, a_p \mathbf{1}_{[K_p]}^{\sf T} ] \}
\end{equation}

where: $\mathbf{1}_{[K_j]}$ is a vector of ones and $K_j$ is the number of quantile-vectors of each block matrix $\mathbf{Q}_j$.

The second step of the MFA consists in a global PCA of the matrices
$\mathbf{Q}_j$ normalized by the  $a_j$ by considering the weights of the individuals that are assumed all constant and equal to $\frac{1}{n}$. The matrix of the weights of the individual is denoted as $\mathbf{W}$.

In such a way, the MFA is equivalent to an analysis on the triplet $(\mathbf {Q,W,A})$ according to the classical definition of the French school (see \cite{Lebart_M}).

The eigensolutions can be obtained by a Generalised SVD of the matrix $\mathbf{Q}$:

\begin{equation}
\mathbf{Q}= \mathbf{U} \Lambda \mathbf{V}^{\sf T}
\end{equation}

under the constraints:

\begin{equation}
\mathbf{U}^T \mathbf{W} \mathbf{U} = \mathbf{V}^{\sf T} \mathbf{A} \mathbf{V} = \mathbf{I}.
\end{equation}

Note that for simplicity the eigenvector and eigenvalues matrices are denoted with same letters than in SVD.

The factor scores of the single quantile vectors of $\mathbf{Q}_j$ are computed as follows:

\begin{equation}
\Psi_{j,\alpha}= a_j \mathbf{Q}_j^{\sf T} \mathbf{v}_{\alpha}
\end{equation}

where $\mathbf{v}_{\alpha}$ is the eigenvector associated with the $\alpha-th$ eigenvalue ($\alpha=1, \ldots, L$ where $L$ is the rank of $\mathbf{S}$).

The factor scores represent a sort of compromise for a common representation in a reduced subspace of the variability structure of the matrices $\mathbf{Q}_j$.

The compromise factor score  $\Psi_{\alpha}$ is the barycenter of the partial factor scores obtained as the average of the $p$ partial scores factors

\begin{equation}
\Psi_{\alpha}= \frac{1}{p} \sum_j a_j \mathbf{Q}_j^T \mathbf{v}_{\alpha}.
\end{equation}

The representation of the individuals (the rows of  $\mathbf{Q_j}$) can be obtained according to the classical biplot on the reduced subspaces as:

\begin{equation}
\Phi_{j,\alpha}= \frac{1}{\lambda_\alpha} a_j \mathbf{Q}_j \mathbf{u}_{\alpha}.
\end{equation}
\vspace{2pt}

\section{TOOLS FOR THE INTERPRETATION: THE SPANISH FAN PLOT}\label{sec:4}

Starting from the results of the MFA, it is interesting the interpretation of the proximities between the distributions according to characteristics that have more contributed to the determination of the axes.

Indeed, in the determination of the factorial axes, the components related to the location, scale and shape, in which the variance (based on the $\ell_2$ Wasserstein distance) of the distributional variable $Y$ can be decomposed, play a different role. This is evident from the results obtained in the next section on simulated data. Even if the analysis is performed on the quantiles, surprisingly, each factorial axis is oriented into the direction of the variability of the means (location parameters), of the standard deviations (size parameters), of the skewness and kurtosis (shape parameters), respectively.
Therefore, the advantage of the proposed approach is to interpret the axes according to the different characteristics of the distribution-valued data. Whatever, if a MFA is performed on sets of four variables (namely, one set for for each distributional variable) representing the first four moments of the distributions, the results are not so evident.

As in classical PCA, the representation of the quantile-variables on the factorial planes (e.g. the first plane for $\alpha$ equal to 1 and 2) is given by in a circle of correlation by the quantile vectors. For improving the interpretation of the plots, we connect the consecutive quantiles  according to their natural order on the factorial plan. It arises a nice representation of the quantile-vectors that remember a \emph{Spanish fan}. We called this representation as  \emph{Spanish-fan plot}. Each \emph{Spanish-fan} allows the analysis of the structure of global variability and the visualization of the characteristics (variability and shape) of the distributional variable.
We observe that the quantile variables representation usually follows a kind of order (being, in general, two consecutive quantile-variables more correlated w.r.t. two non consecutive ones).
We can explain the pattern of the \emph{fan} with respect to  the correlation (i.e., the angle) between pairs of consecutive quantile-variables $q_l$ and $q_{l+k}$ ($k=l+1,\ldots,s$). For example, it is interesting to observe that, when the distributions are almost symmetric the correlation between $q_l$ and $q_{l+k}$ decreases as $k$ ($k=1,2,\ldots,s-l+1$) increases.
Thus, the shape of a \emph{Spanish-fan plot} impacts on the interpretation of the factorial plans. When distribution are different according to their first four moments, we show in the example as the first plane better explains the variability of the locations and  scales of the distributions: the more the \emph{fan} is open, the higher is the variability of the distributions; while the second factorial plane (third and fourth axis) usually explains the variability in skewness and kurtosis of the distributions.

Other typical measures, like the relative contribution, denoted (\emph{cr}), can aid to interpret the axes. Similarly to the classical PCA, the relative contribution of the $i-th$ distribution to the determination of the $\alpha-th$ axis is a measure of how much the variance explained by the $\alpha-th$ axis is due to the $l-th$ quantile variable.

Further, the quality of the representation of the individuals (distributions) and of the quantile variables is measured by the absolute contributes, denoted  (\emph{ca}).
Similarly to classic PCA, absolute contributions sum to one for each distribution (respectively, for each quantile-variable) and the higher is the contribution the better the distribution is represented on the axis (or on the plane, if we consider the sum of the \emph{ca}'s related to the axes of the plane).
\vspace{2pt}

\section{AN APPLICATION OF MFA ON SIMULATED DATA}\label{sec:5}
\noindent
In this section, we present an application of the proposed
MFA on simulated data.
For simplicity, we consider only two distributional variables. The simplicity of the proposed application aims at highlighting the power of the method, especially as visualization tool.

\begin{figure}
	\centering
	\includegraphics[width=0.7\columnwidth]{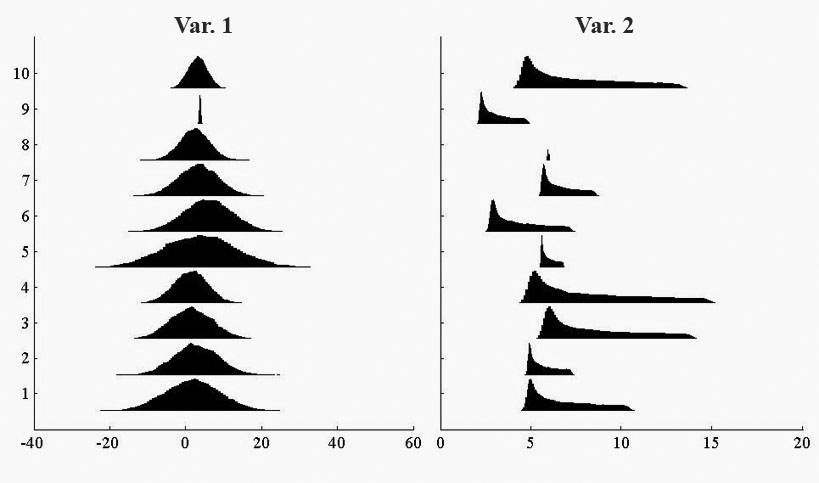}
	\caption{ Representation of the two sets of distributional variables}\label{fig:immagine1}
\end{figure}

\begin{figure}
	\centering
	\includegraphics[width=0.4\columnwidth]{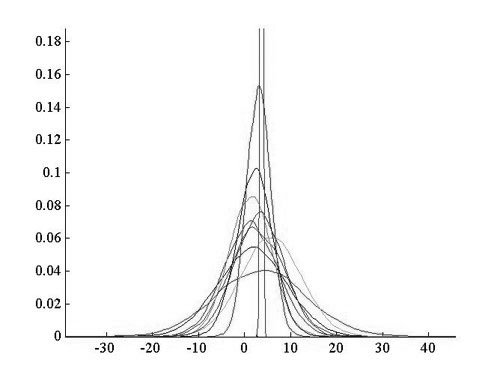}
	\includegraphics[width=0.4\columnwidth]{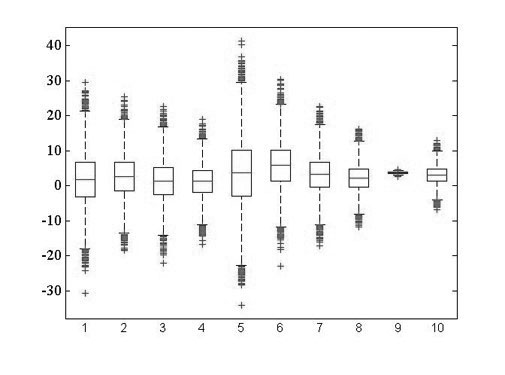}
	\caption{ Characteristics of the point distributions of the first distributional variable $Y_1$} \label{fig:immagine2}
\end{figure}

\begin{figure}
	\centering
	\includegraphics[width=0.4\columnwidth]{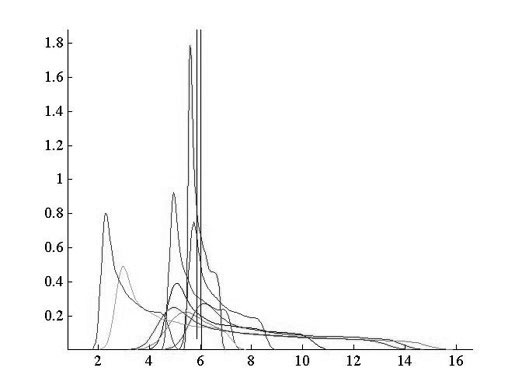}
	\includegraphics[width=0.4\columnwidth]{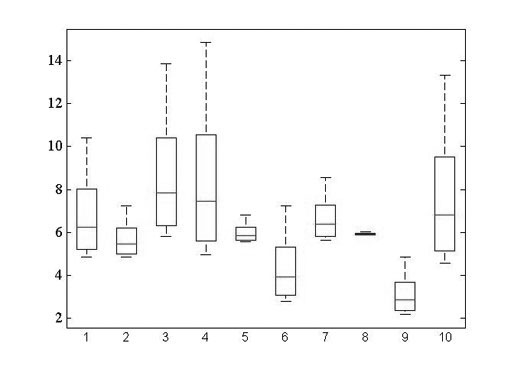}
	\caption{ Characteristics of the point distributions of the second distributional variable $Y_2$}\label{fig:immagine3}
\end{figure}

Considering that the proposed MFA method provides the latent structure of the quantiles for each variable according to the first four moments of the distributions, we considered two set of histogram data observed for the same $n$ individuals. The data related to the first quantile variable are sampled from Gaussian distributions with the same mean and different standard deviations; the second one, are sampled from shifted and scaled Beta distributions. Ten histogram data for each variables have been generated as follows: one thousand of points are sampled for each distribution,
19 quantiles have been extracted, such that each bin, bounded by two consecutive quantiles, contains $5.55\%$ of the sampled points.
In this case, it is equivalent to set up a equi-depth
histogram for each distribution having 18 bins.
The two configurations are shown, using smoothed representations, in Fig. \ref{fig:immagine1}. The box-plots of the sampled data for each distribution are represented in Fig. \ref{fig:immagine2} and in Fig. \ref{fig:immagine3}.

A partial PCA is performed on each block of quantile variables, related to $Y_1$ and $Y_2$.
$\mathbf{Q_1}$ and $\mathbf{Q_2}$ matrices are constituted of 19 quantile variables (including the min values), respectively. The quantile variables are centered w.r.t. the corresponding mean values but they are not scaled.

\begin{figure}
	\centering
	\includegraphics[width=0.4\columnwidth]{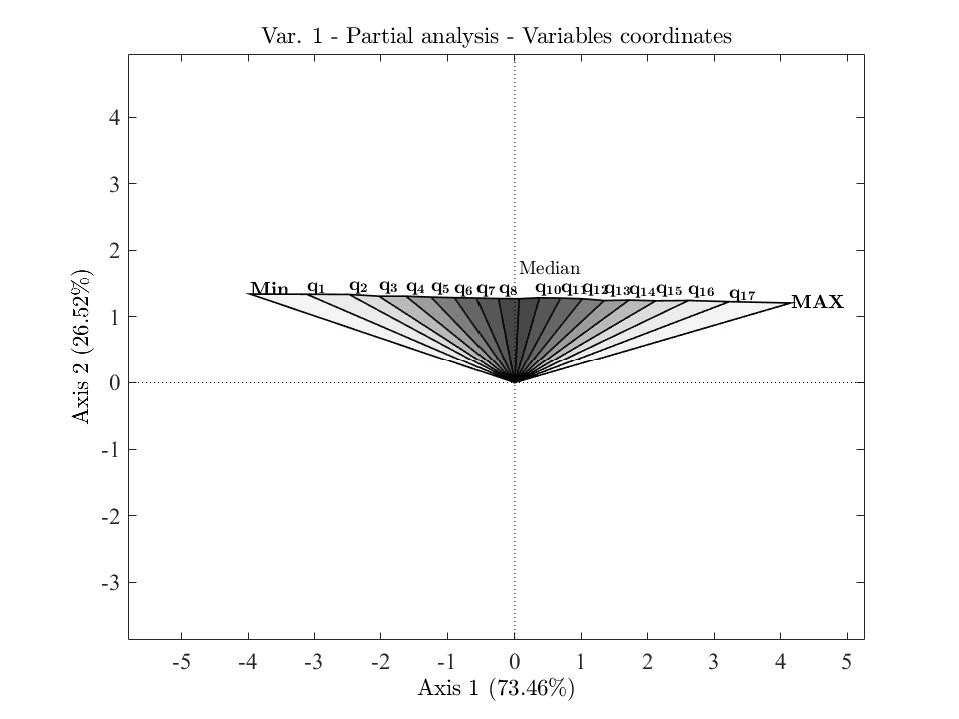}
	\includegraphics[width=0.4\columnwidth]{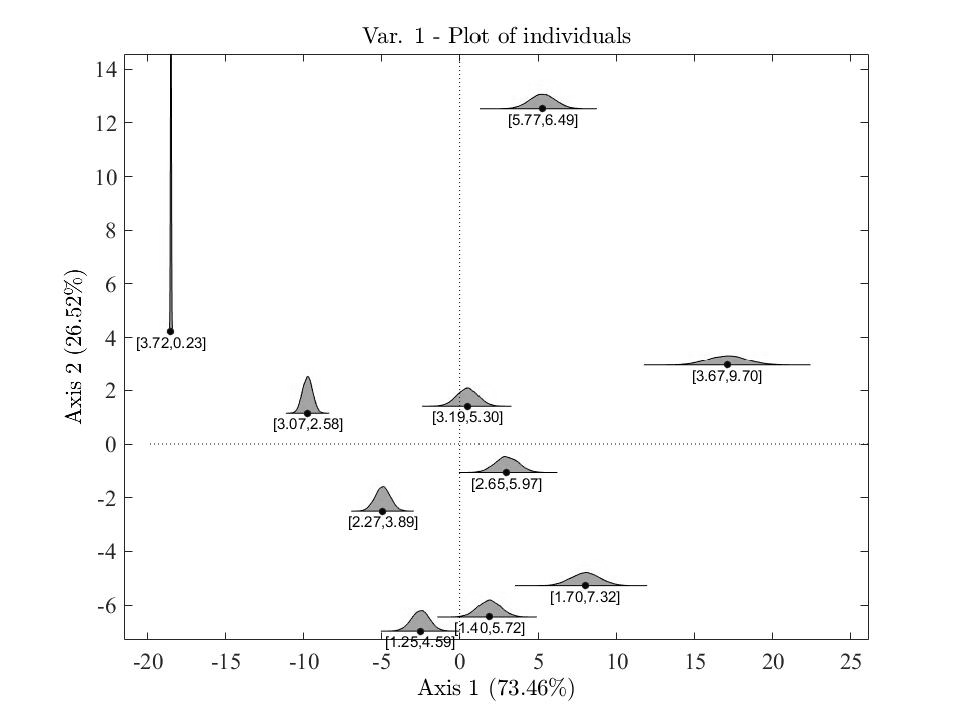}
	\caption{\small Representation of the quantile variables of $Y_1$ on the first factorial plan (\%Explained inertia=99,98\%)} \label{fig:immagine4}
\end{figure}

In this first step, MFA decomposes the Covariance matrices $\mathbf{S_1}$ and $\mathbf{S_2}$, respectively.
Consistently with the characteristics of the first distributional variable $Y_1$, the first latent factor is related to the variability of standard deviations, as we can observe in the two plots in Fig. \ref{fig:immagine4}. Indeed, while the correlation is not so strong with the central quantile-variables because all the Gaussians have the same mean (and median), we note that the first axis is strongly correlated to the extreme quantile-variables.
Each distribution, suitably scaled horizontally and vertically, is overlapped to the point related to the individual such that the mean corresponds to the  abscissa of the point (right panel of Fig. \ref{fig:immagine4}). Following the first axis direction, it is worth noting that the distributions are ordered from the lower to higher value of the std's.

\vspace{2pt}

\begin{figure}
	\centering
	\includegraphics[width=0.4\columnwidth]{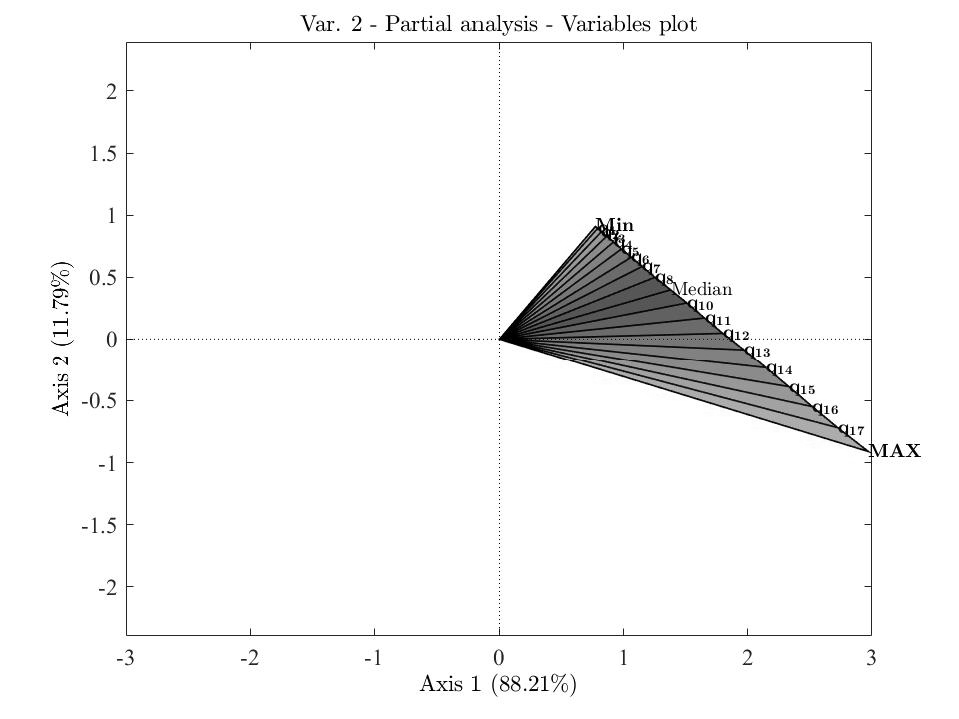}
	\includegraphics[width=0.4\columnwidth]{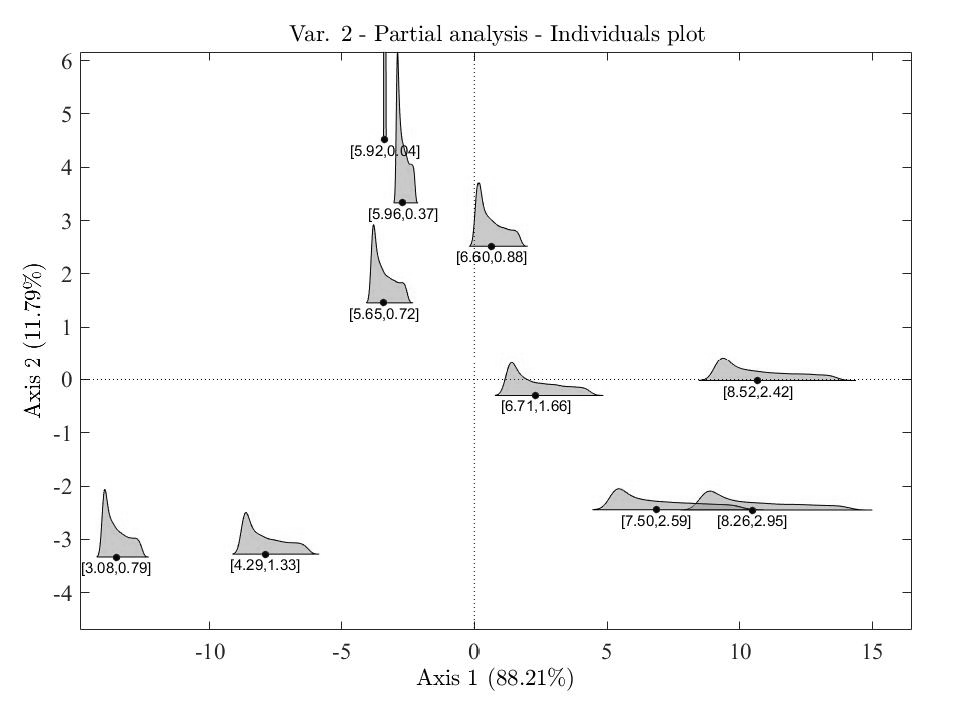}
	\caption{ Representation of the quantile variables of $Y_2$ on the first factorial plan (\%Explained inertia=100\%)} \label{fig:immagine5}
\end{figure}

The partial PCA on the second set of quantile-variables associated with $Y_2$, is based on the decomposition of the Covariance matrix denoted by $\mathbf{S_2}$.
In Fig. \ref{fig:immagine5} are shown the representation of the variables by the \textit{Spanish-fan} plot (on the left) and of the individuals, by overlapping the distributions (on the right) on the first factorial plane.

Since the distributions are all skewed, the representation of the quantile-variables on the first plane (which explains the $100\%$ of the total inertia, with the first axis the $88.21\%$) is very different from the representation of the set of quantile variables associated to $Y_1$. Further, the shape of the \textit{Spanish-fan} (the left panel of Fig. \ref{fig:immagine5}), a scalene triangle, is related to the fact that all the distributions are right skewed. Observing the representation of the individuals by the projected distributions (on the right side of Fig. \ref{fig:immagine5}), it is worth noting that along the first axis the distributions are placed from the lower to the higher mean values (the first value between brackets at the bottom of each distribution) while the second factorial axis opposes the distributions with higher std's values to ones with lower std's values (the second value between brackets).

\begin{figure}
	\centering
	\includegraphics[width=0.52\columnwidth]{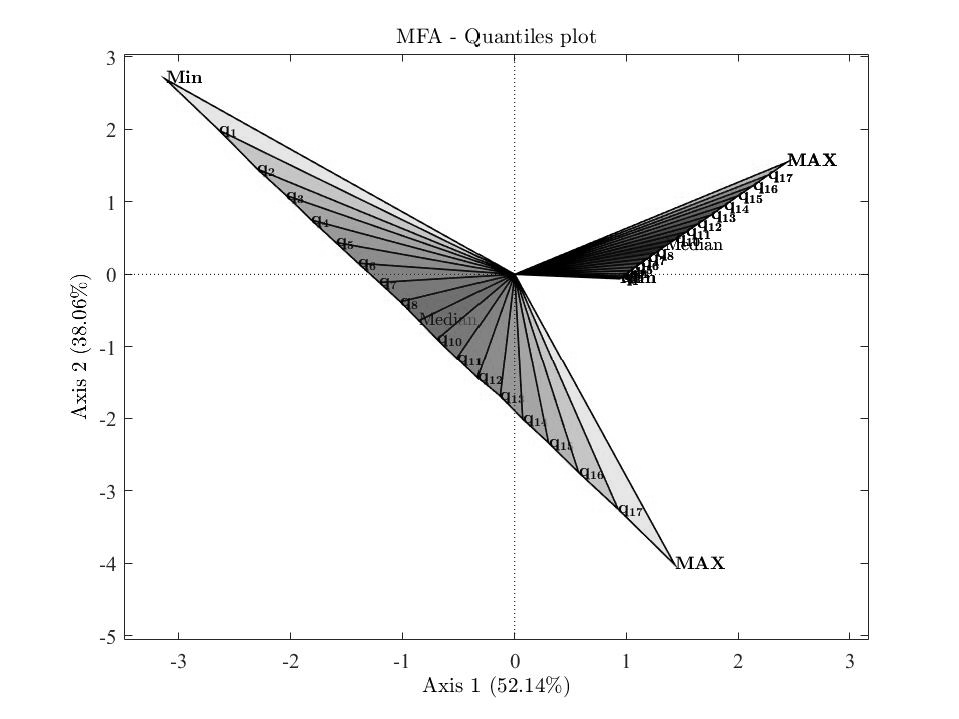}
	\includegraphics[width=0.38\columnwidth]{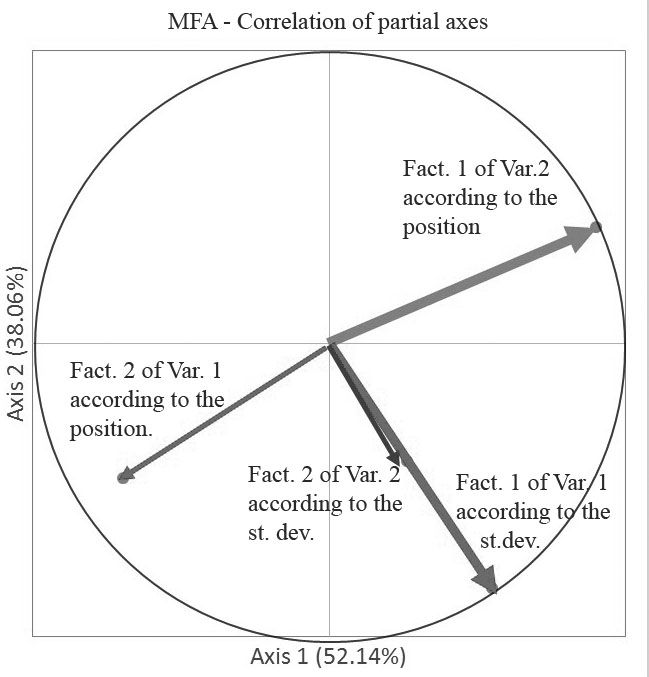}
	\caption{ On the left side: Representation of the quantile variables of $Y_1$ and $Y_2$ on the first factorial plan (\%Explained inertia=90.2\%).
				On the right side: The circle of correlations between the factorial axes of the partial analyses and the factor axes of the global analysis.}   \label{fig:immagine6}
\end{figure}

The second step of the MFA is performed on the global matrix $\mathbf{Q}$. In Fig. \ref{fig:immagine6} is shown a simultaneous representation, on the first factorial plane, of the \textit{Spanish-fans} of the two sets of quantile-variables (on the right side). The explained inertia of the first two factorial axes is $90.2\%$.
Since in the partial analysis, the first \textit{Spanish-fan} of the set of quantile-variables in $\mathbf{Q_1}$ was strongly related to the variability component of the distributions (std), while the second \textit{Spanish-fan} of the set of quantile-variables in $\mathbf{Q_2}$ was characterized by the values of the means on the first axis and to the values of the std on the second axis, in the global analysis, the first \textit{Spanish-fan} plot appears rotated along the second dimension, that is related to the variability of the distributions whereas the first axis is influenced by the values of the means. That is also explained by the graphical representation (on the right side of Fig. \ref{fig:immagine6}) of the correlations between the factorial axes of the partial analyses and the ones of the global analysis.

\begin{figure}
	\centering
	\includegraphics[width=0.9\columnwidth]{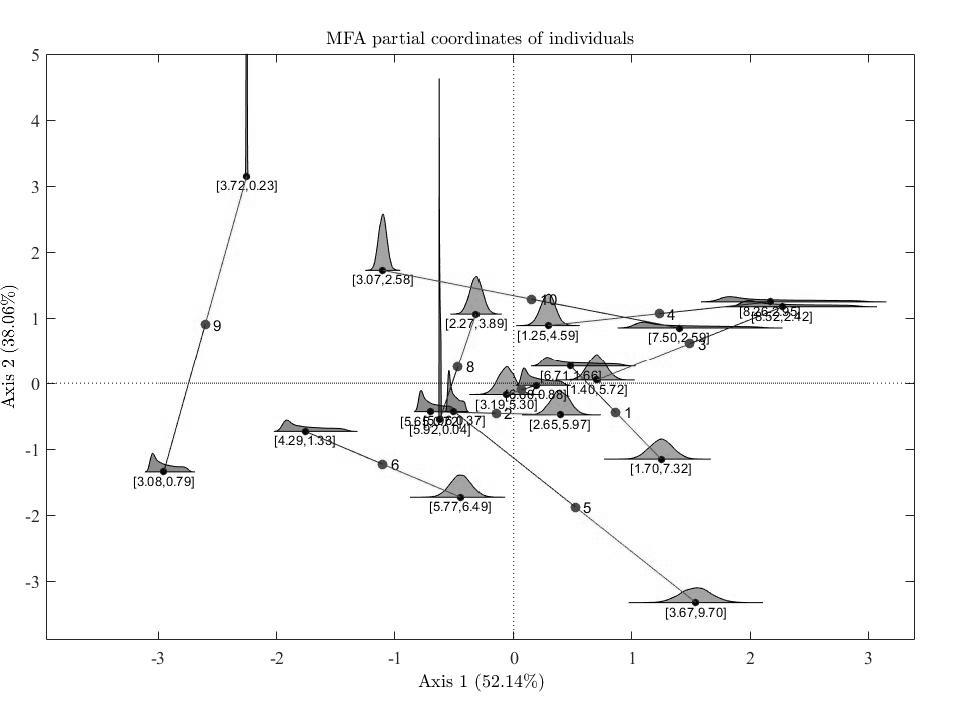}
	\caption{ Representation of the individual distributions with respect to $Y_1$ and $Y_2$ on the first factorial plane (\%Explained inertia=90.2\%). The distributions are drawn on the partial coordinates of individuals while the global coordinates of the individuals are labeled by integers.}\label{fig:immagine9}
\end{figure}

The representation of the individuals on the first factorial plane is displayed in Fig. \ref{fig:immagine9}. The points labeled by numbers are the projection of the individuals on the first plane carried out by the MFA global phase. For interpreting the position of the individual points with respect to the characteristics of the two distributions (empirical realization of $Y_1$ and $Y_2$ distributional variables) for each individual has been projected in supplementary the respective distribution of the variables $Y_1$ and $Y_2$.

\begin{figure}
	\centering
	\includegraphics[width=0.9\columnwidth]{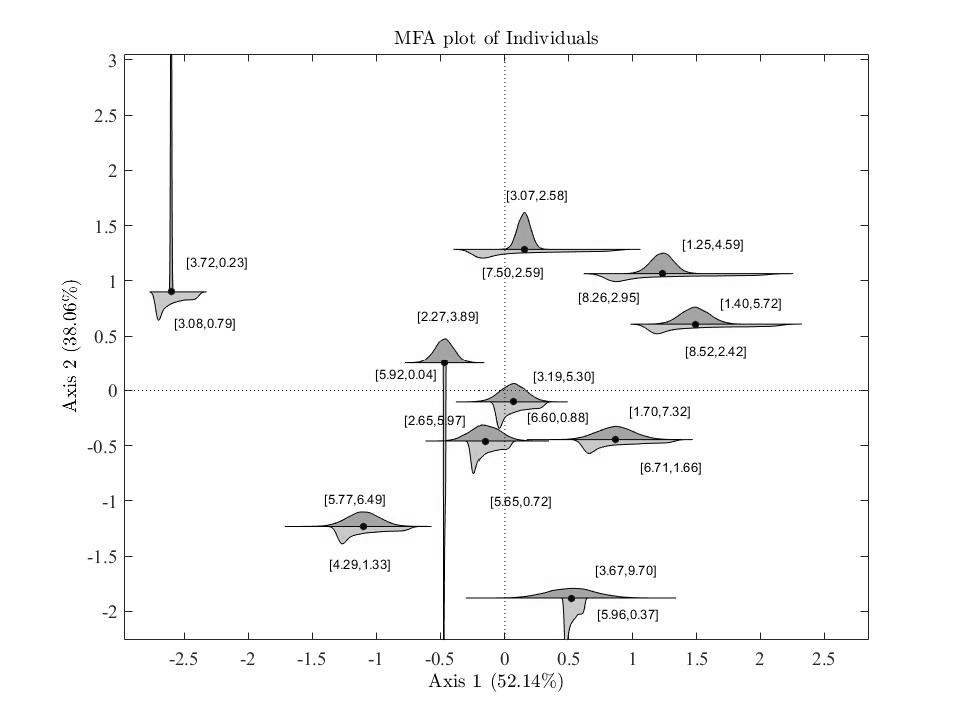}
	\caption{ Representation of the individual distributions with respect to $Y_1$ and $Y_2$ on the first factorial plane (\%Explained inertia=90.2\%). The distributions are drawn on the top and on the bottom of the global coordinates of individuals.}\label{fig:immagine10}
\end{figure}

\begin{figure}
	\centering
	\includegraphics[width=0.9\columnwidth]{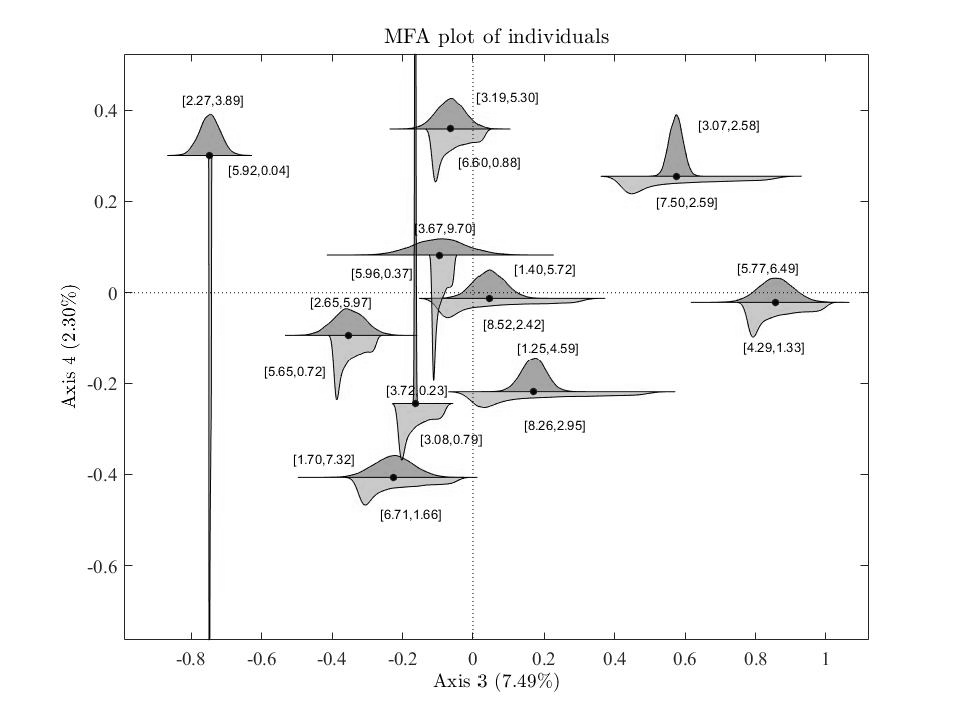}
	\caption{ Representation of the individual distributions with respect to $Y_1$ and $Y_2$ on the second factorial plan (\%Explained inertia=9,79\%). }    \label{fig:immagine11}
	
\end{figure}

In Fig. \ref{fig:immagine10} a different representation of the individual on the first factorial plane is proposed. It is obtained by projecting the distributions of each individual respect to the two variables in the same location points in a specular way. That has been possible since we have just two distributions for each individual. In Fig. \ref{fig:immagine11}, individuals are represented on the factorial plane spanned by the third and fourth axis.

\begin{figure}
	\centering
	\includegraphics[width=0.8\columnwidth]{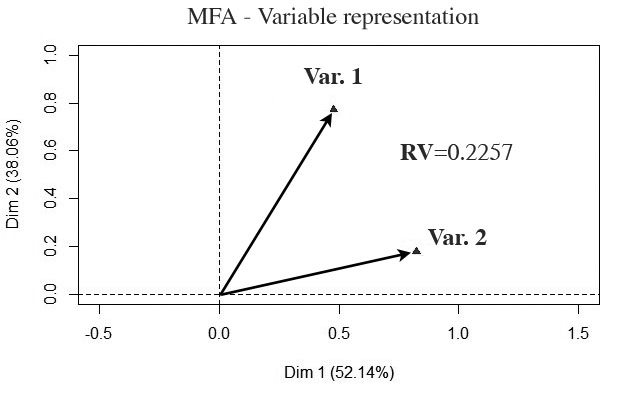}
	\caption{ Correlation between the synthesis of the two sets of quantile-variables expressed by the two vectors Var1 and Var2. RV=0.2257. }    \label{fig:immagine12}
\end{figure}

In Fig. \ref{fig:immagine12} are drawn the vectors corresponding to the two distributional variables. The correlation between the synthesis of the two sets of quantile-variables is expressed by the cosine of the angle on the factorial plane according to the classical measure RV proposed by \cite{MFA83_1}, that in this case is $RV=0.2257$.

\vspace{2pt}

\section{AN APPLICATION OF MFA ON REAL DATA}\label{sec:6}
\begin{figure}
	\centering
	\includegraphics[width=0.9\columnwidth]{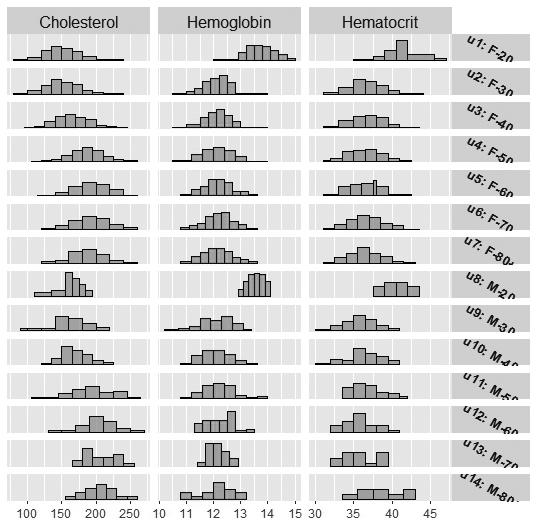}
	\caption{ The BLOOD dataset }    \label{fig:immagine20}
\end{figure}

\noindent
In this section, we present an application of the proposed
MFA on a dataset described in \cite{BillDid06}. The dataset is the description of the \textit{Cholesterol}, \textit{Hemoglobin} and \textit{Hematocrit} levels observed for 14 groups of patients (each group is identified by a sex-age typology), using histograms of values. The size and the raw data of each group is not available, thus a classical PCA is not possible. The dataset is also available in the \texttt{HistDAWass} package\footnote{\url{https://cran.r-project.org/web/packages/HistDAWass/index.html}} developed in R. The data table is shown in Fig. \ref{fig:immagine20}.
The analysis is performed using twenty quantiles for each histogram. The MFA returns components with associated eigenvalues as presented in Tab. \ref{Tab:eigs_BLOOD_MFA}. We note that the first two components synthesize the $96.16\%$ in the the total variance, thus we represent the main results using only the first factorial plane.

\begin{table}[ht]
	\centering
	\caption{ BLOOD dataset: eigenvalues of each component}
	\begin{center}
		\begin{widetable}{0.9\columnwidth}%
			{@{\extracolsep{\fill}\vrule}*4{ c|}}
			\hline
			Components & Eigenvalue & \% of variance & cum. \% of variance \\
			\hline
			comp 1 & 2.28 & 71.55 & 71.55 \\
			comp 2 & 0.67 & 20.98 & 92.53 \\
			comp 3 & 0.12 & 3.63 & 96.16 \\
			comp 4 & 0.07 & 2.12 & 98.27 \\
			comp 5 & 0.02 & 0.58 & 98.85 \\
			comp 6 & 0.02 & 0.48 & 99.33 \\
			comp 7 & 0.01 & 0.25 & 99.59 \\
			comp 8 & 0.01 & 0.18 & 99.76 \\
			comp 9 & 0.00 & 0.13 & 99.89 \\
			comp 10 & 0.00 & 0.06 & 99.95 \\
			comp 11 & 0.00 & 0.03 & 99.98 \\
			comp 12 & 0.00 & 0.01 & 99.99 \\
			comp 13 & 0.00 & 0.01 & 100.00 \\
			\hline
		\end{widetable}
	\end{center}
	\label{Tab:eigs_BLOOD_MFA}
\end{table}
\subsection{Representation of distributional variables}
\begin{figure}
	\centering
	\includegraphics[width=0.7\columnwidth]{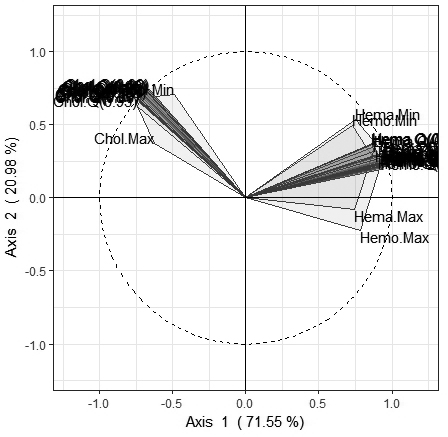}
	\caption{ MFA first factorial plane: Spanish fan plots}    \label{fig:immagine21}
\end{figure}

\begin{figure}
	\centering
	\includegraphics[width=0.7\columnwidth]{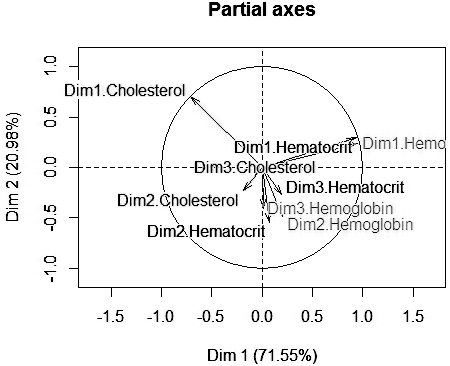}
	\caption{ MFA first factorial plane: Correlation among the first three dimensions of the partial PCA for each distributional variable}    \label{fig:immagine22}
\end{figure}

The representation of variables is performed by means of the Spanish fan plots. Since each set of quantiles define a block of variables in the MFA, we show the correlation plot of the Spanish fans on the first factorial plane. In Fig. \ref{fig:immagine21} we have the plot of the Spanish fans, while in Fig. \ref{fig:immagine22}, it is possible to see the correlation of the dimensions of each partial PCA for each variable.

In Fig. \ref{fig:immagine21}, we note that \textit{Hemoglobin} and \textit{Hematocrit} fans are quite overlapping, while the \textit{Cholesterol} fan is perpendicular, on average, with respect the others. Further, observing the span of the funs, it appears that distributions related to \textit{Cholesterol} are less variable in scale (otherwise the span should be more open) than those related to \textit{Hemoglobin} and \textit{Hematocrit}.
In Fig. \ref{fig:immagine22}, we observe that the first dimensions of the partial analysis for the  \textit{Hemoglobin} and \textit{Hematocrit} are positively correlated, while their correlation with the first dimension of \textit{Cholesterol} is very weak. Further the second dimension are very low on the first factorial plane. Looking at both the figures, we note that the first dimensions of the three distributional variables follow the direction of the central quantiles, thus, they are mainly related to the variability of positions. The other dimensions, that are related to the variability of scales and shapes are very short, thus, these aspects play very little in explaining the variability of the distributional data.

\subsection{Representation of individuals}
\begin{figure}
	\centering
	\includegraphics[width=0.45\columnwidth]{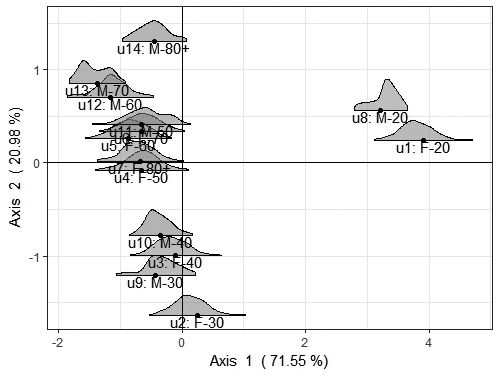}
	\includegraphics[width=0.45\columnwidth]{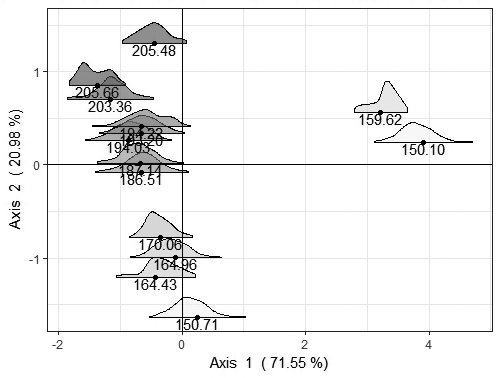}
	\caption{ MFA first factorial plane: Plots of individuals for \textit{Cholesterol} variable, on the left data are labeled with the object name, on the right data are labeled with the mean value.}   \label{fig:immagine23}
\end{figure}

\begin{figure}
	\centering
	\includegraphics[width=0.45\columnwidth]{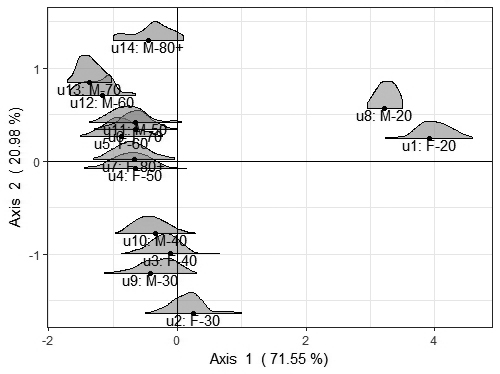}
	\includegraphics[width=0.45\columnwidth]{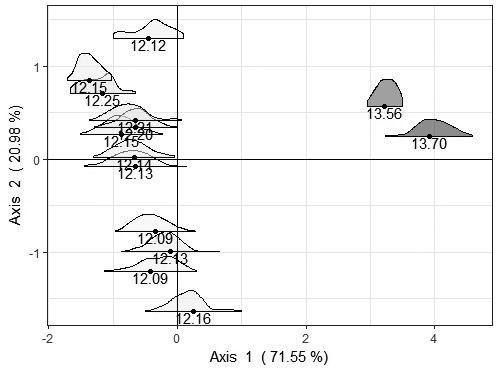}
	\caption{MFA first factorial plane: Plots of individuals for \textit{Hemoglobin} variable, on the left data are labeled with the object name, on the right data are labeled with the mean value.}   \label{fig:immagine24}
\end{figure}

\begin{figure}
	\centering
	\includegraphics[width=0.45\columnwidth]{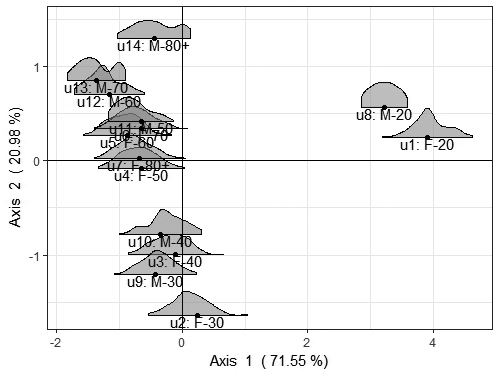}
	\includegraphics[width=0.45\columnwidth]{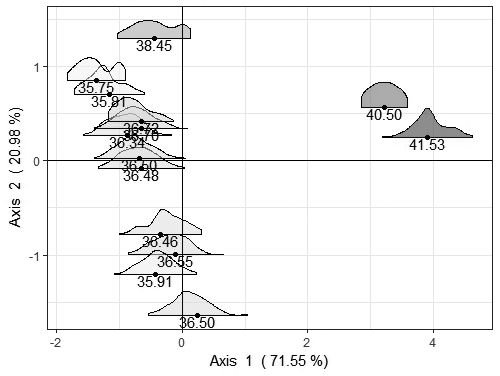}
	\caption{ MFA first factorial plane: Plots of individuals for \textit{Hematocrit} variable, on the left data are labeled with the object name, on the right data are labeled with the mean value.}   \label{fig:immagine25}
\end{figure}

The representation of individuals on the first factorial plane is performed by juxtaposing the original distribution on the coordinates of each individual. Each distribution is arranged such that the mean of each distribution corresponds to the coordinates of each individual. For showing the main characteristics of the objects according their distribution for each variables we show in Figs. \ref{fig:immagine23}, \ref{fig:immagine24} and \ref{fig:immagine25} the distributions for, respectively, the \textit{Cholesterol}, \textit{Hemoglobin} and \textit{Hematocrit} variable. Each plot is organized such that on the left, individuals are labeled according their name, while, on the right, individuals are labeled according their mean value and the darker the distributions are, the higher is their mean. In this way, we observe that, for each distributional variable, the direction of the increase of the means is consistent with the direction of the first dimension for each variable as shown in Fig. \ref{fig:immagine22}. Further, we may compare the scale and the shape of each distribution.


\subsection{Comments}
As expected, being them related to the quantity of iron in the blood cells,  the \textit{Hemoglobin}  and \textit{Hematocrit} are positively correlated and their values are higher in young people.  \textit{Cholesterol} has a low correlation with the other two variable, and the its mean value tends to increase from younger to older peoples. About the other scale and shape comparisons there are very slight differences between distributions thus, the analysis is few influenced from those aspects. It is possible to compare the scale and the shape of \textit{M-20} and \textit{F-20} on the factorial planes for the  \textit{Hemoglobin}  and \textit{Hematocrit} variable observing that at the top of the plane there are distributions that are generally less variable and with a low kurtosis (Tab. \ref{Tab_kurt} than those at the bottom.
\begin{table}[ht]
	\centering
	\caption{ BLOOD dataset: kurtosis of each distribution computed as the fourth standardized moment.}
	\begin{center}
		\begin{widetable}{0.9\columnwidth}%
			{@{\extracolsep{\fill}\vrule}*4{ c|}}
			\hline
			Objects& Cholesterol & Hemoglobin & Hematocrit \\
			\hline
			u1: F-20 & 3.37 & 3.08 & 3.00 \\
			u2: F-30 & 3.23 & 4.10 & 2.72 \\
			u3: F-40 & 2.95 & 3.60 & 2.50 \\
			u4: F-50 & 3.18 & 2.81 & 2.51 \\
			u5: F-60 & 2.65 & 2.77 & 2.58 \\
			u6: F-70 & 2.68 & 2.91 & 2.61 \\
			u7: F-80+ & 3.04 & 2.72 & 2.73 \\
			u8: M-20 & 2.85 & 2.14 & 1.96 \\
			u9: M-30 & 3.16 & 2.83 & 2.73 \\
			u10: M-40 & 2.83 & 2.61 & 2.66 \\
			u11: M-50 & 2.74 & 3.44 & 2.49 \\
			u12: M-60 & 3.37 & 2.34 & 2.63 \\
			u13: M-70 & 1.92 & 2.37 & 1.86 \\
			u14: M-80+ & 2.56 & 2.42 & 1.94 \\
			\hline
		\end{widetable}
	\end{center}
	\label{Tab_kurt}
\end{table}

\section{CONLUSIONS}\label{sec:conl}
\noindent
This paper represents an extension of the MFA for the definition of a PCA method for distributional data based on the $\ell_2$ Wasserstein distance between distributions. We showed that the trace of the covariance matrix of the quantile-variables approximates the variance of a distributional variable computed with the Wasserstein metric. Previous approaches was not related to a particular metric between distributions and, thus, a comparison could not be appropriate.
Using quantile-variables, we observed that the proposed PCA is able to identify the differences in the structure of the several sets of variables in the analysis according to the main characteristics of the distributional variables: position, scale and shape. The classical MFA on standard data is here enriched by the nature of the analyzed data. The characteristics of the observed distributions are here emphasized by the peculiar tools for the interpretation. Further, a novel \emph{Spanish-fan} plot has been introduced for describing the relations among the quantile-variables projected on the factorial planes. We showed how to interpret the shape of a fan with respect to the characteristics of the distributions. Therefore, the similarity between the distributions (individuals in the analysis) is well interpreted according to the similarity between their parameters on each axis. The proposed applications on simulated and real data have shown how each axis is strongly related to the variability of parameters of position, scale and shape, respectively.
Aiming at showing the advantages of the method and give more readable factorial planes, only few distributional variables have been considered in the applications.
	
	
	
	
	\section*{References}
	\bibliographystyle{model1a-num-names}
	\bibliography{references}
	
	
	
	
	
	

\end{document}